\documentclass[conference]{IEEEtran}
\IEEEoverridecommandlockouts
\usepackage[numbers]{natbib}
\usepackage{amsmath,amssymb,amsfonts}
\usepackage{algorithmic}
\usepackage{graphicx}
\usepackage{float}
\usepackage{algorithm}
\usepackage{hyperref}

\ifCLASSOPTIONcompsoc
    \usepackage[caption=false, font=normalsize, labelfont=sf, textfont=sf]{subfig}
\else
\usepackage[caption=false, font=footnotesize]{subfig}
\fi
\usepackage{textcomp}
\usepackage[table]{xcolor} % Add this package with the 'table' option
\usepackage{tabularx} % Add this package
\usepackage{booktabs} % Optional, for a cleaner table appearance
\usepackage{colortbl}

\usepackage{tikz}
\usetikzlibrary{shapes,arrows,positioning,calc}

\definecolor{headercolor}{RGB}{40, 110, 160}
\definecolor{rowcolor1}{RGB}{215, 240, 255}
\definecolor{rowcolor2}{RGB}{180, 220, 240}
\definecolor{white}{RGB}{255, 255, 255}

\def\BibTeX{{\rm B\kern-.05em{\sc i\kern-.025em b}\kern-.08em
    T\kern-.1667em\lower.7ex\hbox{E}\kern-.125emX}}
\begin{document}

\title{Rafting Towards Consensus: Formation Control of Distributed Dynamical Systems\\
}
% 1\textsuperscript{st}
\author{\IEEEauthorblockN{Abbas Tariverdi\IEEEauthorrefmark{1},
		 Jim Tørresen\IEEEauthorrefmark{2}}\\
	\IEEEauthorblockA{\IEEEauthorrefmark{1}Department of Physics,
		University of Oslo, Oslo, Norway\\
		\IEEEauthorrefmark{2}Department of Informatics,
		University of Oslo, Oslo, Norway\\
		\thanks{
			Corresponding author: Abbas Tariverdi (email: abbast@uio.no).}}}

\maketitle

\begin{abstract}
    % In this work, we present a novel adaptation of the Raft Consensus Algorithm to achieve emergent formation control in multi-agent systems with single integrator dynamics. Dubbed "Rafting," this approach enables robust cooperation among disparate nodes, facilitating the attainment of desired geometric configurations. Our framework capitalizes on the inherent fault tolerance and strong consistency guarantees of the Raft algorithm, extending its applicability to distributed formation control tasks. We introduce a decentralized mechanism for aggregating agent states, followed by a synchronization protocol for information exchange and consensus formation. The Raft consensus algorithm elegantly fuses leader election, log replication, and state machine application to drive agents towards a shared objective. A series of meticulous simulations validate the efficacy and resilience of our approach under varying conditions, including partial network failures, communication delays, and perturbations. The results demonstrate the potential of the presented algorithm, opening new horizons in swarm robotics, autonomous transportation, and distributed sensing.
    In this paper, we introduce a novel adaptation of the Raft consensus algorithm for achieving emergent formation control in multi-agent systems with a single integrator dynamics. This strategy, dubbed "Rafting," enables robust cooperation between distributed nodes, thereby facilitating the achievement of desired geometric configurations. Our framework takes advantage of the Raft algorithm's inherent fault tolerance and strong consistency guarantees to extend its applicability to distributed formation control tasks. Following the introduction of a decentralized mechanism for aggregating agent states, a synchronization protocol for information exchange and consensus formation is proposed. The Raft consensus algorithm combines leader election, log replication, and state machine application to steer agents toward a common, collaborative goal. A series of detailed simulations validate the efficacy and robustness of our method under various conditions, including partial network failures and disturbances. The outcomes demonstrate the algorithm's potential and open up new possibilities in swarm robotics, autonomous transportation, and distributed computation. The implementation of the algorithms presented in this paper is available at \href{https://github.com/abbas-tari/raft.git}{https://github.com/abbas-tari/raft.git}.
\end{abstract}

\begin{IEEEkeywords}
Distributed Dynamical Systems, Multi-Agent Systems, Raft, Consensus, Formation Control, Fault Tolerant
\end{IEEEkeywords}

\section{Introduction}
Distributed Dynamical Systems (DDSs) consist of multiple interacting agents collaborating to achieve complex goals that exceed the capabilities of individual agents \cite{wooldridge2009introduction}. Distributed systems allow tasks to be divided across spatially dispersed agents, enhancing robustness, scalability, and effective resource allocation \cite{coulouris2005distributed}. In recent decades, multi-agent systems (MAS) and distributed systems have risen to prominence in cooperative tasks, such as formation control \cite{hu2020distributed} and collaborative computation \cite{kempe2003gossip}. Formation control enables agents to achieve predetermined geometric arrangements, which is vital for applications in swarm robotics \cite{brambilla2013swarm}, Unmanned Aerial Vehicle (UAV) networks \cite{he2016distributed}, and sensor arrays \cite{martinson2005lattice}. 

% Collaborative computation leverages distributed processing, allowing the system to solve intricate problems more efficiently and reliably \cite{kempe2003gossip}. These paradigms continue to reshape modern computing, communication, and control systems.

By leveraging the potential of MAS and distributed systems, new solutions for a wide variety of applications, including traffic management, smart grids, distributed data mining, and disaster response can be devised \cite{cao2013overview}. 
% The fusion of multi-agent and distributed systems has not only facilitated real-time decision-making and reduced computational complexity but also improved the adaptability and reliability of large-scale networks. 
The importance of MAS and distributed systems will only increase as technology advances and the demand for more efficient, cooperative systems rises, leading to further developments in the field \cite{stone2000multi}.

% The Raft consensus algorithm, proposed by Ongaro and Ousterhout \cite{ongaro2014search}, is a distributed consensus protocol designed for managing replicated logs in fault-tolerant systems. 
For handling replicated logs in fault-tolerant systems, Ongaro and Ousterhout \cite{ongaro2014search} introduced a distributed consensus technique called Raft.
It distinguishes itself from the conventional Paxos algorithm \cite{lamport1998part} by providing simplicity, readability, and robust consistency guarantees.
Raft is based on a leader-follower paradigm, employing leader election, log replication, and state machine application to maintain consistency across nodes \cite{howard2014raft}.  
% Its applications span distributed databases, configuration management, and coordination services. 
Raft plays a crucial role in cooperative tasks and distributed systems by guaranteeing that system components perform coherently and consistently, despite failures or delays. Its properties make it suited for allowing robust multi-agent collaboration across a variety of areas, from IoT networks to distributed computing systems \cite{moraru2013more}.

Utilizing the Raft consensus algorithm for formation control in dynamical multi-agent systems presents a novel approach to address coordination challenges. The leader-follower paradigm of Raft ensures fault tolerance and strong consistency, critical for MAS reliability \cite{ongaro2014search}. By adapting Raft to single integrator MAS, the algorithm can promote resilient collaboration across spatially separated agents, enabling them to achieve desired geometric configurations despite network failures or communication delays \cite{olfati2004consensus}. 

This paper is organized into five sections to describe the  application of the Raft consensus algorithm for formation control in MAS. The Background section provides a thorough  introduction to the Raft consensus algorithm and formation control in MAS, detailing their fundamentals, characteristics, and relevance to the problem domain. The Method section elaborates on the employment of the Raft algorithm for formation control, outlining the integration of single integrator dynamics with Raft's inherent mechanisms. In the Simulation section, we present a series of carefully designed simulations that validate the efficacy and resilience of our proposed approach under varying conditions. The Discussion section analyzes the results, highlighting the advantages and potential challenges of the presented approach. Finally, the Conclusion section summarizes the key contributions of our study, while also exploring future research directions to advance the field.

\section{Background}
The Raft consensus algorithm, introduced by Ongaro and Ousterhout \cite{ongaro2014search}, revolutionized the landscape of distributed consensus protocols by providing a more understandable and accessible alternative to the Paxos algorithm \cite{lamport1998part}. Since its inception, numerous studies have been conducted to evaluate, extend, and apply the Raft algorithm in various domains.

One of the earliest evaluations of the Raft consensus algorithm was presented by Howard et al. \cite{howard2014raft}. The authors investigated its key characteristics, assessed its performance, and compared it with other consensus algorithms. Their study contributed to the validation and acceptance of Raft in the distributed systems community.
Moraru et al. \cite{moraru2013more} extended Raft to Egalitarian parliaments, a more generalized consensus model. Their work, "There is more consensus in Egalitarian parliaments," presented a novel approach to consensus decision-making and demonstrated the versatility of the Raft algorithm.

%Jin et al. \cite{jin2018optimized} focused on improving the performance and scalability of Raft in distributed databases. Their study, "An optimized consensus algorithm based on the Raft protocol," proposed optimization techniques that significantly enhanced the efficiency of Raft-based systems.
%Sundaravel et al. \cite{sundaravel2018raft} applied the Raft consensus algorithm in IoT networks to address the challenges of scalability and reliability. Their paper, "Raft-based consensus for large-scale IoT networks," showcased the potential of Raft in handling the unique requirements of IoT environments.

Fu et al. \cite{fu2021improved} optimizes the Raft consensus algorithm for the Hyperledger Fabric platform in terms of both log replication and leader election.
Liu et al. \cite{liu2021dqn} proposes a new leader selection scheme based on the distributed consensus algorithm RAFT to facilitate the use of blockchain on resource-constrained IoT end devices.
Wang et al. \cite{wang2019k} discusses a Raft-like consensus algorithm that preserves the logic of part of Raft consensus algorithm and optimizes leader election and consensus process through the established K-Bucket node relationships in the Kademlia protocol and demonstrated the applicability of Raft in the burgeoning field of blockchain technology.

Overall, these studies exemplify the adaptability and robustness of the Raft consensus algorithm across diverse applications. Each study contributed to the understanding and advancement of Raft, highlighting its potential for enabling reliable and efficient distributed systems.

Table \ref{table:raft_comparison} summarizes and compares a few recent articles and researches on the Raft consensus algorithm:

\begin{table}[H]
    \centering
    \caption{Comparison of Raft-related work}
    \label{table:raft_comparison}
    \begin{tabularx}{\linewidth}{>{\columncolor{white}}X|>{\columncolor{white}}X|>{\columncolor{white}}X}
    \toprule 
    \rowcolor{headercolor}
    \textcolor{white}{\textbf{Reference}} & \textcolor{white}{\textbf{Key Contributions}} & \textcolor{white}{\textbf{Domain/Application}} \\ \midrule
    \rowcolor{rowcolor1}
    Ongaro and Howard et al.\cite{ongaro2014search, howard2014raft} & Introduced and Evaluated Raft & Distributed systems \\ \midrule
%    \rowcolor{rowcolor2}
%    Howard et al.\cite{howard2014raft} & Evaluated Raft & Distributed systems \\ \midrule
    \rowcolor{rowcolor2}
    Moraru et al.\cite{moraru2013more} & Extended Raft & Egalitarian parliaments \\ \midrule
    \rowcolor{rowcolor1}
    Liu et al.\cite{liu2021dqn} & Applied Raft & IoT networks \\ \midrule
    \rowcolor{rowcolor2}
    Wang and Fu et al.\cite{wang2019k,fu2021improved} & Used Raft & Blockchain systems \\ \midrule
    \rowcolor{rowcolor1}
    Gonzalez et al.\cite{gonzalez2016fault} & Raft for  Software-Defined Networking (SDN) & Software-defined networking \\ \midrule
%    \rowcolor{rowcolor2}
%    Sharma et al.\cite{sharma2018craq} & Raft with CRDTs & Distributed data structures \\ \midrule
    \rowcolor{rowcolor2}
	Deyerl et al.\cite{deyerl2019search} & Scalable Raft & Distributed systems \\ \midrule
%    \rowcolor{rowcolor1}
%    Ardekani et al.\cite{ardekani2014samsara} & Raft in geo-replication & Cloud computing \\ \midrule
    \rowcolor{rowcolor1}
    Hou et al.\cite{hou2021intelligent} & Raft for edge computing & IoT and Edge computing \\ \midrule
    \rowcolor{rowcolor2}
    Nasimi et al.\cite{nasimi2020platoon} & Raft in vehicular networks & Vehicular Ad-Hoc networks (VANETs) \\ 

%\midrule    
%    \rowcolor{rowcolor2}
%    Liu et al.\cite{liu2020raft} & Raft for UAVs & Unmanned aerial vehicles \\ 
    \bottomrule
\end{tabularx}
\end{table}

Formation control and consensus in dynamical multi-agent systems have gained significant attention over the past decade due to their wide applicability in various domains, including robotics, distributed computing, and swarm intelligence. This literature review highlights ten recent studies that have contributed to the field in unique ways.

One study by Chen et al. \cite{cheng2019distributed} studies the formation control problem for general linear multi-agent systems constrained with unavailable states and event-triggered communications. Li et al. \cite{li2013leader} considers the leader-following formation control problem for second-order multiagent systems with time-varying delay and nonlinear dynamics. Xiao et al. \cite{xiao2007consensus} explore the consensus problem in high-dimensional systems, presenting a scalable solution that can be adapted to different applications.

In the area of robotic swarms, Kartal et al. \cite{kartal2020distributed}proposes a backstepping-based, distributed formation control method that is stable independent of time delays in communication among multiple unmanned aerial vehicles (UAVs).  

For practical applications, Nair et al. \cite{nair2018multi} presents a comprehensive literature review on the application of Multi-agent systems (MAS) to Economic Dispatch (ED) and Unit Commitment (UC) in smart grids, ensuring efficient and reliable operation. Qian et al. \cite{qian2023formation} apply formation control techniques to mobile sensor networks for environmental monitoring, using a consensus-based approach to improve the sensing quality.

In the realm of connected vehicles, Wang et al. \cite{wang2018review} reviews researches regarding different aspects of Cooperative adaptive cruise control (CACC) systems. CACC is one promising technology to allow connected and automated vehicles (CAVs) to be driven in a cooperative manner and introduces system-wide benefits.. Liu et al. \cite{liu2016cooperative} proposes a novel robust cubature Kalman filter (CKF) for cooperative localization of connected vehicles under a GNSS/DSRC integrated architecture. The CKF is enhanced using the Huber M-estimation technique to improve the performance of data fusion under uncertain sensor observation environments. Finally, in the context of social network analysis, Karimi et al. \cite{karimi2018consensus} introduce proposes a consensus-based community detection approach (CBC) for finding communities in multilayer networks.

\begin{table}
    \centering
    \caption{Summary of key contributions and domain/applications of formation control studies}
    \label{tab:formation_control_studies}
    \begin{tabularx}{\linewidth}{>{\columncolor{white}}c|>{\columncolor{white}}X|>{\columncolor{white}}X}
        \toprule
        \rowcolor{headercolor}
        \textcolor{white}{\textbf{Reference}} & \textcolor{white}{\textbf{Key Contributions}} & \textcolor{white}{\textbf{Domain/Application}} \\
        \midrule
        \rowcolor{rowcolor1}
        Cheng et al.\cite{cheng2019distributed} & Distributed formation control via output feedback event-triggered coordination & Robotics, multi-agent systems \\ \midrule
        \rowcolor{rowcolor2}
        Li et al.\cite{li2013leader} & Leader-following formation control for second-order multiagent systems with time-varying delay and nonlinear dynamics & Multi-agent systems \\ \midrule
        \rowcolor{rowcolor1}
        Xiao et al.\cite{xiao2007consensus} & Consensus in high-dimensional multi-agent systems & Distributed computing, swarm intelligence \\ \midrule
        \rowcolor{rowcolor2}
        Kartal et al.\cite{kartal2020distributed} & Distributed formation control for UAV swarms with communication delays & UAV swarms, robotics \\ \midrule
%        \rowcolor{rowcolor1}
%        Liu et al.\cite{liu2017consensus} & Consensus-based approach for target tracking using quadrotor swarm & Robotics, UAV swarms \\ \midrule
        \rowcolor{rowcolor1}
        Nair et al.\cite{nair2018multi} & Distributed consensus algorithm for smart grid applications & Smart grids, energy management \\ \midrule
        \rowcolor{rowcolor2}
        Qian et al.\cite{qian2023formation} & Formation control for mobile sensor networks using consensus-based approach & Environmental monitoring, sensor networks \\ \midrule
        \rowcolor{rowcolor1}
        Wang et al.\cite{wang2018review} & Formation control for cooperative adaptive cruise control & Connected vehicles, traffic management \\ \midrule
        \rowcolor{rowcolor2}
        Liu et al.\cite{liu2016cooperative} & Distributed consensus algorithm for cooperative localization in connected vehicle systems & Connected vehicles, localization \\ \midrule
        \rowcolor{rowcolor1}
        Karimi et al.\cite{karimi2018consensus} & Consensus-based methodology for detection communities in multilayered networks &  Network analysis, community detection \\ 
        \bottomrule
    \end{tabularx}
\end{table}

\section{Problem Formulation and Main Results}
The Raft consensus algorithm is typically used for achieving consensus in distributed systems, particularly in applications like distributed databases and configuration management.

Applying Raft to dynamical systems such as robots can be beneficial in certain scenarios. For example, if you have a group of robots that need to coordinate their actions and maintain a consistent state, Raft could help ensure that all robots agree on the state of the system and the actions to be taken. This can be particularly useful in environments where communication is unreliable, or where there is a risk of individual robot failures.

In a formation control scenario, the Raft consensus algorithm allows agents to communicate with each other through a graph, share their states, and adjust their behavior based on the received information. In other words, the Raft algorithm manages the agents' states as the replicated log entries and updates and synchronizes the states across all nodes. Whenever an agent updates its state, the Raft algorithm would be used to replicate the new state across all nodes. The agents would then adjust their behavior based on the received state information.

The Raft consensus algorithm is not used to solve the formation control problem in a traditional sense. Instead, it is used to synchronize the states (i.e., positions) of the agents across different nodes in a distributed system.

We assume that each agent is on a different node in the Raft algorithm. In this case, each agent will be managed by a separate node in the Raft cluster, and these nodes will use the Raft algorithm to reach a consensus on the agents' states.

When each agent is on a separate node, it means that each agent's state is managed independently within its associated node. The Raft algorithm is responsible for ensuring that the agent's state is consistent across all nodes. This will allow the agents to cooperate and reach the desired formation on a regular polygon.

Let's consider a group of $N$ agents, and let their positions in a 2D plane be denoted as $x_i = (x_i, y_i)$ for $i = 1, 2, \ldots, N$. The desired formation can be represented as a formation graph $G = (V, E)$, where $V$ is the set of vertices (each vertex represents an agent), and $E$ is the set of edges (each edge represents the desired relationship between two agents).

The formation control problem can be posed as a consensus problem:

Define the relative position vector $d_{ij}$ for each pair of connected agents $(i, j)$ in the graph $G$ as:
\[d_{ij} = x_j - x_i\]

Define the desired relative position vector $d_{ij}^*$ for each pair of connected agents $(i, j)$ in the graph $G$ as:
\[d_{ij}^* = x_j^* - x_i^*\]

where $(x_i^*, y_i^*)$ represents the desired position of agent $i$ in the formation.

Define the formation error $e_{ij}$ for each pair of connected agents $(i, j)$ in the graph $G$ as:
\[e_{ij} = d_{ij} - d_{ij}^*\]

The formation control objective is to design a control law $u_i$ for each agent $i$ such that:
\[u_i = -k \sum_{\substack{j=1\\ j\neq i}}^N L_{ij} e_{ij}\]

where $k$ is a positive control gain, and $L_{ij}$ is the $ij$-th element of the Laplacian matrix $L$ of the formation graph $G$. The Laplacian matrix $L$ can be computed as $L = D - A$, where $D$ is the degree matrix (a diagonal matrix with the degree of each vertex on the diagonal) and $A$ is the adjacency matrix (with $A_{ij} = 1$ if $(i, j)$ is an edge in $E$ and $A_{ij} = 0$ otherwise).

The dynamics of agent $i$ can be described by a set of differential equations:
\[\frac{dx_i}{dt} = u_i\]

The overall formation control problem can be expressed as finding a control law $u_i$ for each agent $i$ that minimizes the global formation error:
\[E = 0.5 \sum_{(i, j) \in E} e_{ij}^2\]

The control laws designed with the above approach will ensure that the formation error converges to zero, meaning the agents will maintain the desired formation while moving. The choice of control gain $k$ and the structure of the formation graph $G$ determine the performance and stability of the formation control system.

Integration of the Raft communication steps:

\begin{enumerate}
	\item Agent $i$ updates its position $\boldsymbol{x}_i$ and/or control input $\boldsymbol{u}_i$, and sends the update to the Raft leader.
	
	\item The Raft leader appends the update to its log and replicates the log to all followers.
	
	\item Followers receive the log, apply updates to their state machines, and send acknowledgments back to the leader.
	
	\item The leader ensures a majority of agents have applied the update before considering it committed.
\end{enumerate}

The Raft algorithm mainly deals with the communication and state replication process, while the formation control uses the equations to compute the control inputs and maintain the desired formation. By combining the communication infrastructure provided by Raft with the formation control equations, the multi-agent system can achieve the desired formation while ensuring fault-tolerance and consistency in the presence of communication or agent failures.

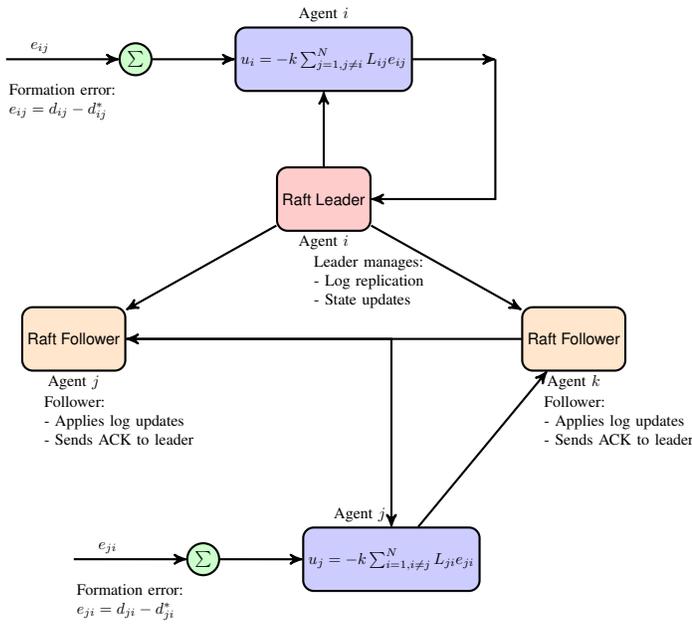
\begin{figure}
	\centering
	\begin{tikzpicture}[auto, node distance=1.1cm,>=latex', scale=0.6, every node/.style={scale=0.6}]
		\tikzset{
			block/.style={draw, thick, rounded corners, rectangle, minimum height=4em, minimum width=4em, font=\sffamily, fill=blue!20},
			sum/.style={draw, thick, circle, inner sep=2pt, node distance=1.5cm, font=\sffamily, fill=green!20},
			arrow/.style={->, >=stealth', thick},
			input/.style={coordinate},
			output/.style={coordinate},
			pinstyle/.style={pin edge={to-,thin,black}}
		}
		
		% Formation control blocks
		\node [input, name=input1] {};
		\node [sum, right = of input1] (sum1) {$\sum$};
		\node [block, right = of sum1] (controller1) {$u_i = -k \sum_{j=1, j\neq i}^N L_{ij} e_{ij}$};
		\node [output, right = of controller1, name=output1] {};
		
		\draw [arrow] (input1) -- node[name=e1, xshift=-0.5cm] {$e_{ij}$} (sum1);
		\draw [arrow] (sum1) -- (controller1);
		\draw [arrow] (controller1) -- (output1);
		
		% Raft blocks
		\node [block, below = 1cm of controller1, fill=red!20] (raft1) {Raft Leader};
		\node [block, below left = 1cm and 2cm of raft1, fill=orange!20] (raft2) {Raft Follower};
		\node [block, below right = 1cm and 2cm of raft1, fill=orange!20] (raft3) {Raft Follower};

		\draw [arrow] (output1) |- (raft1);
		\draw [arrow] (raft1) -- (controller1);
		\draw [arrow] (raft1) -- (raft2);
		\draw [arrow] (raft1) -- (raft3);
		
		% Agent dynamics blocks
		\node [input, below = 2.5cm of raft2] (input2) {};
		\node [sum, right = of input2] (sum2) {$\sum$};
		\node [block, right = of sum2] (controller2) {$u_j = -k \sum_{i=1, i\neq j}^N L_{ji} e_{ji}$};
		\node [output, right = of controller2, name=output2] {};
		
		\draw [arrow] (input2) -- node[name=e2, xshift=-0.5cm] {$e_{ji}$} (sum2);
		\draw [arrow] (sum2) -- (controller2);
%		\draw [arrow] (controller2) -- (output2);
		
		\draw [arrow] (controller2) -- (raft3);
		\draw [arrow] (raft3) -- (raft2);
		\draw [arrow] (raft2) -| (controller2);
		
		% Labels
		\node [above, align=center] at 	(controller1.north) {Agent $i$};
		\node [above, align=center, xshift=-0.7cm] at (controller2.north) {Agent $j$};
		\node [align=center, yshift=-0.25cm] at (raft1.south) {Agent $i$};
		\node [align=center, yshift=-0.25cm] at (raft2.south) {Agent $j$};
		\node [align=center, yshift=-0.25cm] at (raft3.south) {Agent $k$};

		% Technical details
		\node [below = 0.25cm of e1, align=left, xshift=0.5cm] {Formation error:\\ $e_{ij} = d_{ij} - d_{ij}^*$};
		\node [below = 0.25cm of e2, align=left, xshift=0.5cm] {Formation error:\\ $e_{ji} = d_{ji} - d_{ji}^*$};
		\node [below = 0.25cm of raft1, align=left, xshift=1cm] {Leader manages:\\- Log replication\\- State updates};
		\node [below = 0.25cm of raft2, align=left, xshift=1cm] {Follower:\\- Applies log updates\\- Sends ACK to leader};
		\node [below = 0.25cm of raft3, align=left, xshift=1cm] {Follower:\\- Applies log updates\\- Sends ACK to leader};
	\end{tikzpicture}
\caption{Integration of Raft consensus algorithm with formation control of multi-agent systems.}
\label{fig:raft_formation_control}
\end{figure}

\section{Simulations}

For the simulation, we assume that the center of regular polygon is at origin and agents' positions are initialized randomly. One of the agents will be selected as the leader, and it will be responsible for calculating the next positions for all agents, including itself. The other agents will follow the leader's instructions. Throught Scenario \ref{v4EE} to \ref{v5E}, the leader changes periodically every 20 frames. So the original Raft's leader election mechanism is not used.

The agents collaborate by following the updates provided by the leader, and the leader is responsible for maintaining the formation. The targets (goal positions) for each agent are set based on the formation shape, which is determined by the leader. So, the agents are working together to maintain the formation, and their targets are set according to the overall objective of the system, which is to maintain the desired formation.

\subsection{Scenario A}\label{v4EE}
%v4_EE
When the current leader gets disconnected, a new leader is elected automatically by the Raft algorithm to ensure the continued operation of the system. In this process, nodes can vote for a leader based on their unique identifier (address). This way, you can ensure that a new leader will be elected if the current leader gets disconnected.

To demonstrate the effects of changing the leader or handling leader failure, the leader is periodically switched or upon the failure of a leader, agents vote to select the new leader.

Below is an example of periodically switching the leader, the leader node is switched every 20 frames, and a leader failure is simulated at frame 35. 
Then a new leader is elected by incrementing the  failed leader index. In this example it should be noted that the leader failure simulation only prevented the failed leader from updating the positions of other agents, but the failed leader's position was still updated by the new leader.
% FIXME: Why agent 0 is being counted twice!!!! needs to be fixed
%\begin{figure}[htbp]
%    \centering
%  \subfloat[a\label{1a}]{%
%       \includegraphics[width=0.475\linewidth]{Imgs_f_a=n_v4_E/Figure_1.png}}
%    \hfill
%  \subfloat[b\label{1b}]{%
%        \includegraphics[width=0.475\linewidth]{Imgs_f_a=n_v4_E/agent_0.png}}
%    \\
%  \subfloat[c\label{1c}]{%
%        \includegraphics[width=0.475\linewidth]{Imgs_f_a=n_v4_E/agent_1.png}}
%    \hfill
%  \subfloat[d\label{1d}]{%
%        \includegraphics[width=0.475\linewidth]{Imgs_f_a=n_v4_E/agent_2.png}}
%  \caption{Scenario I}
%  \label{fig1} 
%\end{figure}

\begin{figure}[htbp]
	\centering
	\subfloat[a\label{4EEa}]{%
		\includegraphics[width=0.475\linewidth, trim={0pt 0pt 10mm 8mm}, clip]{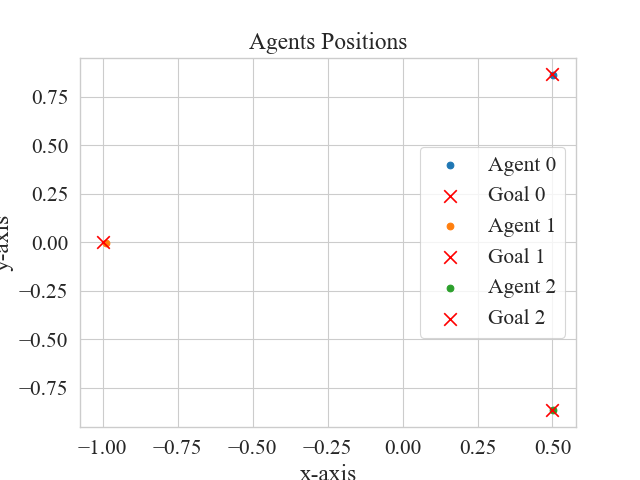}}
	\hfill
	\subfloat[b\label{4EEb}]{%
		\includegraphics[width=0.475\linewidth, trim={7mm 7mm 7mm 40mm}, clip]{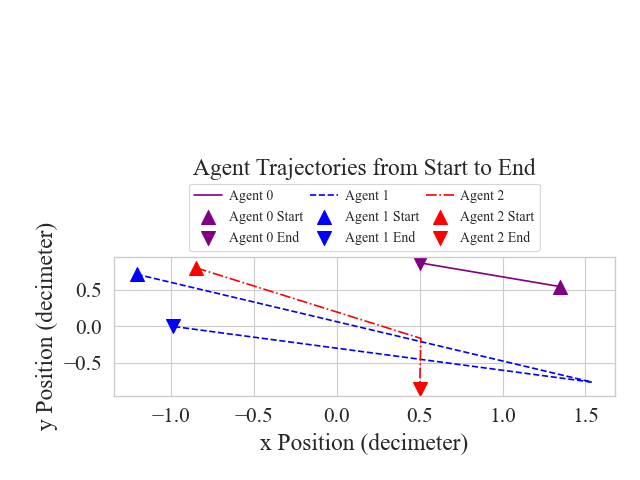}}
	\\
	\subfloat[c\label{4EEc}]{%
		\includegraphics[width=0.475\linewidth, trim={0mm 0mm 7mm 7mm}, clip]{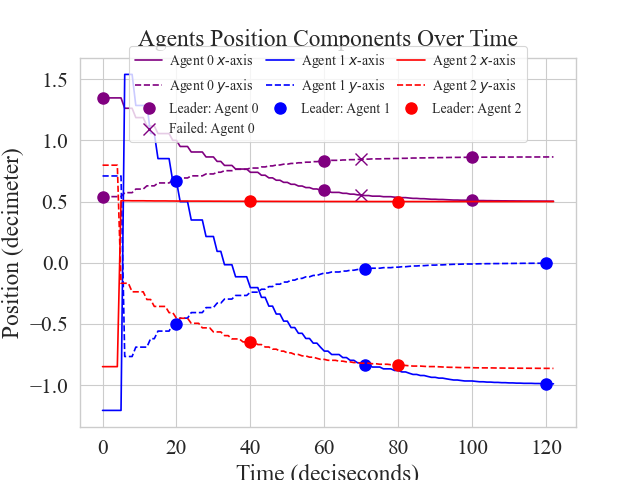}}
	\hfill
	\subfloat[d\label{4EEd}]{%
		\includegraphics[width=0.475\linewidth, trim={3mm 0mm 7mm 7mm}, clip]{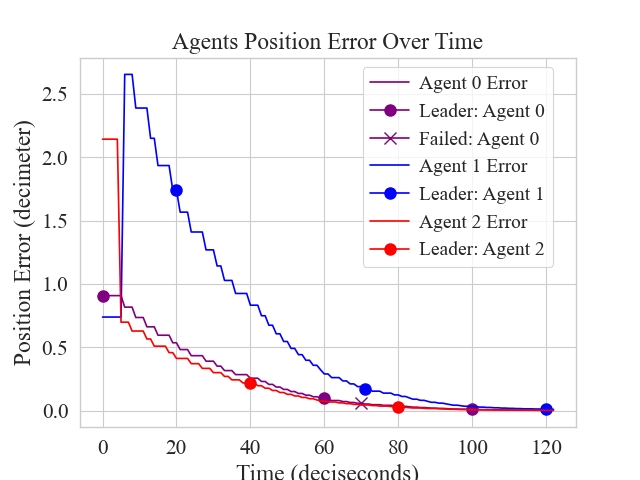}}
	\caption{Scenario A: Leader is periodically switched or upon the failure of a leader, agents vote to select the new leader. Leader node is switched every 20 frames, and a leader failure is simulated at frame 35. Then a new leader is elected by incrementing the  failed leader index. (a): Agents' final position. (b): Agents trajectories from start to end. (c): Agents poistion components ($x$ and $y$) over time. (d): Agents position errors over time.}
	\label{fig4EE} 
\end{figure}

\subsection{Scenario B}\label{v5}
\begin{figure}[htbp]
	\centering
	\subfloat[a\label{5a}]{%
		\includegraphics[width=0.475\linewidth, trim={0pt 0pt 10mm 8mm}, clip]{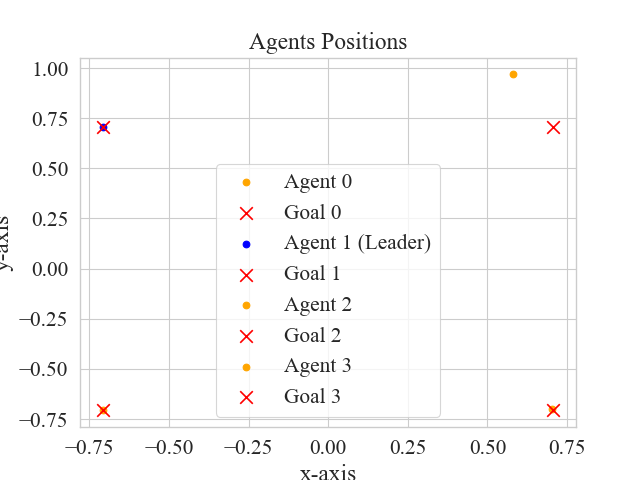}}
	\hfill
	\subfloat[b\label{5b}]{%
		\includegraphics[width=0.475\linewidth, trim={7mm 7mm 7mm 40mm}, clip]{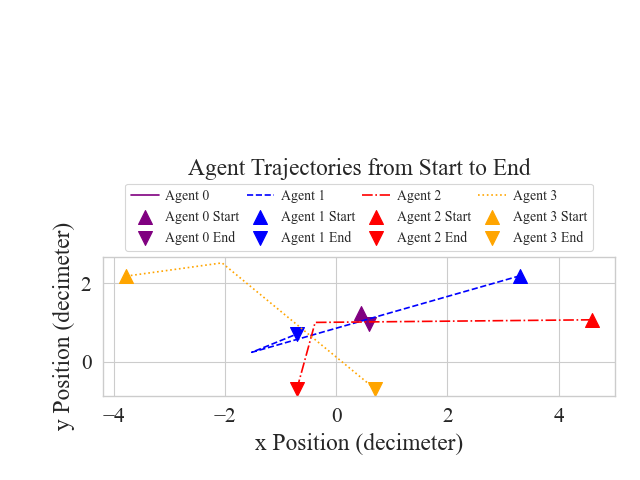}}
	\\
	\subfloat[c\label{5c}]{%
		\includegraphics[width=0.475\linewidth, trim={0mm 0mm 7mm 7mm}, clip]{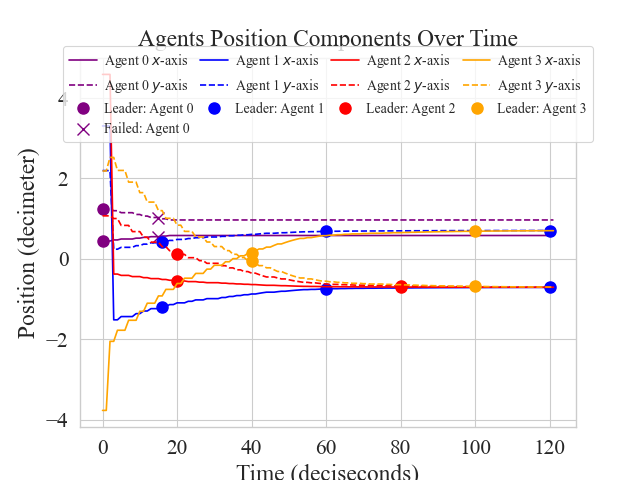}}
	\hfill
	\subfloat[d\label{5d}]{%
		\includegraphics[width=0.475\linewidth, trim={3mm 0mm 7mm 7mm}, clip]{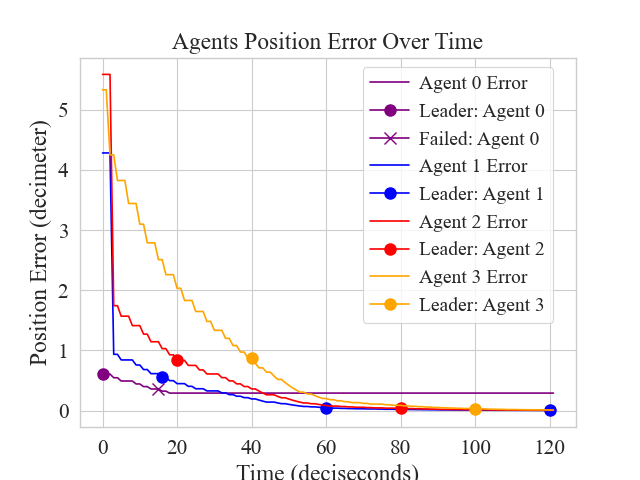}}
	\caption{Scenario B:  When an agent or nodes fails as a leader at a specific time, the position of the failed leader should not be updated anymore showing that the agent is disconnected from the rest. Agnet 1 will be staionary after its failure. (a): Agents' final position. (b): Agents trajectories from start to end. (c): Agents poistion components ($x$ and $y$) over time. (d): Agents position errors over time.}
	\label{fig5} 
\end{figure}

%v5
In this scenario, when an agent or nodes fails as a leader at a specific time, the position of the failed leader should not be updated anymore showing that the agent is disconnected from the rest.
% The issue with this approach is that when an agent fails we have formation on regular polygon order n, however it should be regular polygon order n-1
The other agents make a formation on a regular $n$-sided polygon even if some agents fail and get disconnected from the rest. Figure \ref{fig5} shows that Agnet 1 will be staionary after its failure.

%\begin{figure}[htbp] 
%    \centering
%  \subfloat[a\label{2a}]{%
%       \includegraphics[width=0.475\linewidth]{Imgs_f_a=n_v5/Figure_1.png}}
%    \hfill
%  \subfloat[b\label{2b}]{%
%        \includegraphics[width=0.475\linewidth]{Imgs_f_a=n_v5/agent_0.png}}
%    \\
%  \subfloat[c\label{2c}]{%
%        \includegraphics[width=0.475\linewidth]{Imgs_f_a=n_v5/agent_1.png}}
%    \hfill
%  \subfloat[d\label{2d}]{%
%        \includegraphics[width=0.475\linewidth]{Imgs_f_a=n_v5/agent_2.png}}
%    \\
%    \subfloat[c\label{2e}]{%
%            \includegraphics[width=0.475\linewidth]{Imgs_f_a=n_v5/agent_3.png}}
%  \caption{Scenario B}
%  \label{fig45} 
%\end{figure}

\subsection{Scenario C}\label{v5E}
\begin{figure}[htbp]
	\centering
	\subfloat[a\label{5Ea}]{%
		\includegraphics[width=0.475\linewidth, trim={0pt 0pt 10mm 8mm}, clip]{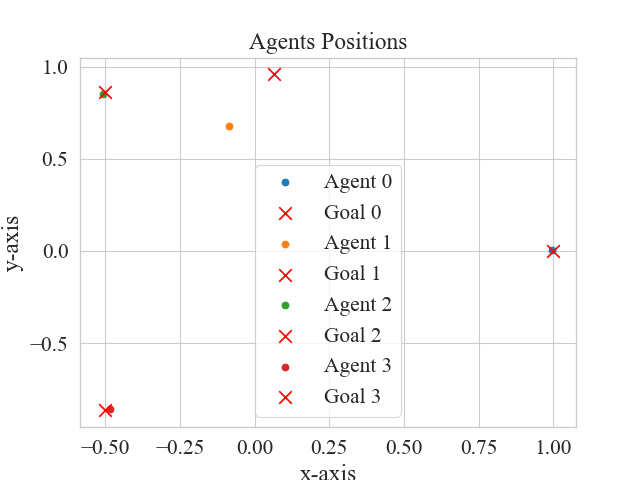}}
	\hfill
	\subfloat[b\label{5Eb}]{%
		\includegraphics[width=0.475\linewidth, trim={7mm 7mm 7mm 40mm}, clip]{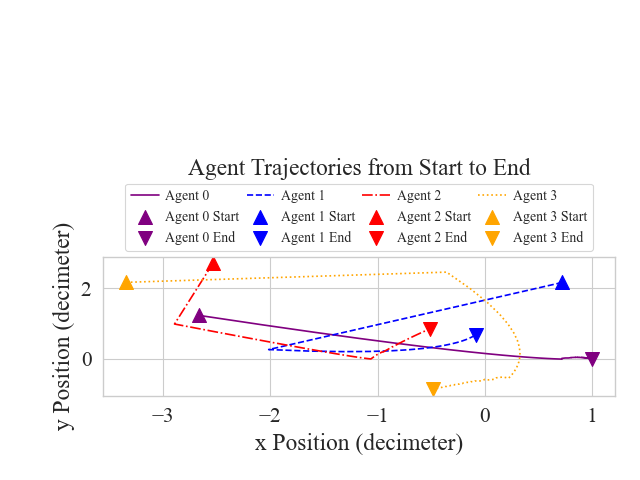}}
	\\
	\subfloat[c\label{5Ec}]{%
		\includegraphics[width=0.475\linewidth, trim={0mm 0mm 7mm 7mm}, clip]{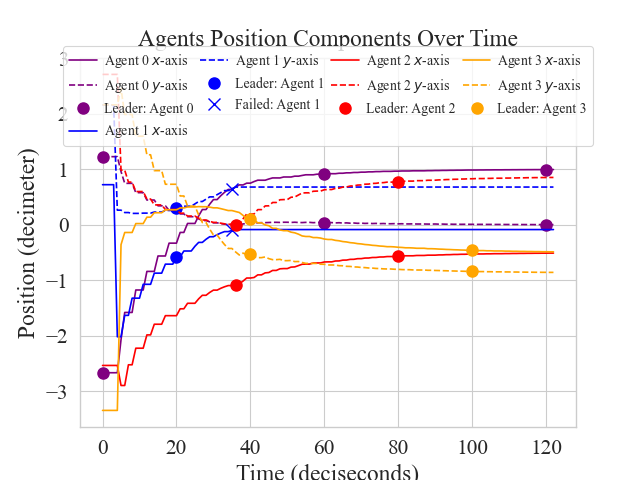}}
	\hfill
	\subfloat[d\label{5Ed}]{%
		\includegraphics[width=0.475\linewidth, trim={3mm 0mm 7mm 7mm}, clip]{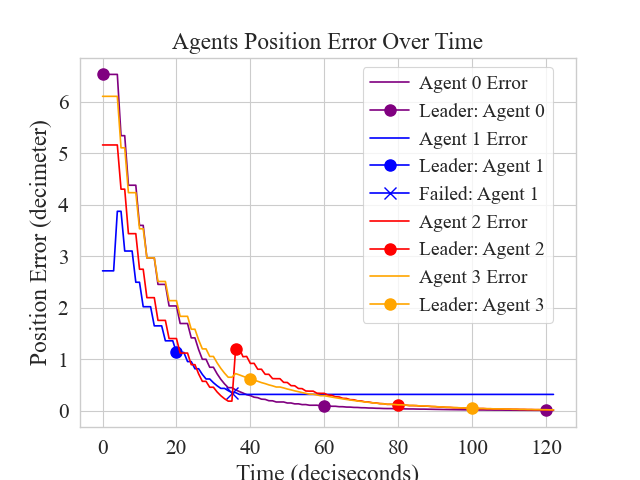}}
	\caption{Scenario C: When an agent or nodes fails as a leader at a specific time, the position of the failed leader is not updated anymore showing that the agent is disconnected from the rest. Other agents make a formation on a regular $n-m$-sided polygon where $m$ is number of fail agents. (a): Agents' final position. (b): Agents trajectories from start to end. (c): Agents poistion components ($x$ and $y$) over time. (d): Agents position errors over time.}
	\label{fig5E} 
\end{figure}

%v5E
In this scenario, when an agent or nodes fails as a leader at a specific time, the position of the failed leader should not be updated anymore showing that the agent is disconnected from the rest.
The other agents make a formation on a regular $n-m$-sided polygon where $m$ is number of fail agents (Figure \ref{fig5E}).

\subsection{Scenario D}\label{v6}
% f_a=n_v6.py
The Raft algorithm ensures that all nodes in the system maintain a consistent state by replicating logs across all nodes. 
This feature can be leveraged to maintain a consistent view of the agent positions and goals across all nodes. Instead of updating the positions only on the leader node and then broadcasting the updated positions to the other nodes, the leader node can propose position updates and let the Raft algorithm replicate the new state across all nodes.

\begin{figure}[htbp]
	\centering
	\subfloat[a\label{6a}]{%
		\includegraphics[width=0.475\linewidth, trim={0pt 0pt 10mm 8mm}, clip]{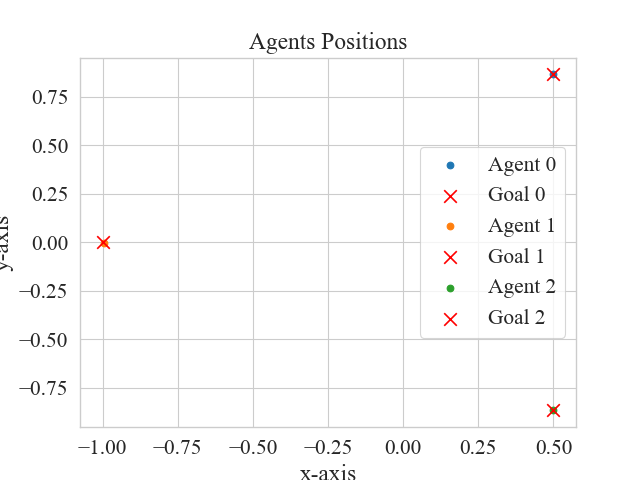}}
	\hfill
	\subfloat[b\label{6b}]{%
		\includegraphics[width=0.475\linewidth, trim={7mm 7mm 7mm 40mm}, clip]{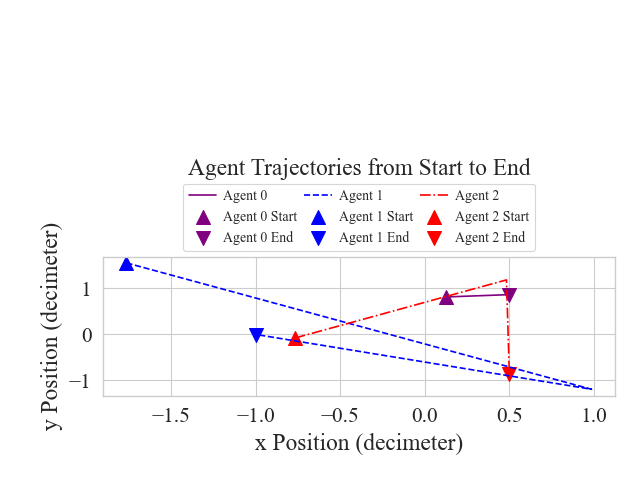}}
	\\
	\subfloat[c\label{6c}]{%
		\includegraphics[width=0.475\linewidth, trim={0mm 0mm 7mm 7mm}, clip]{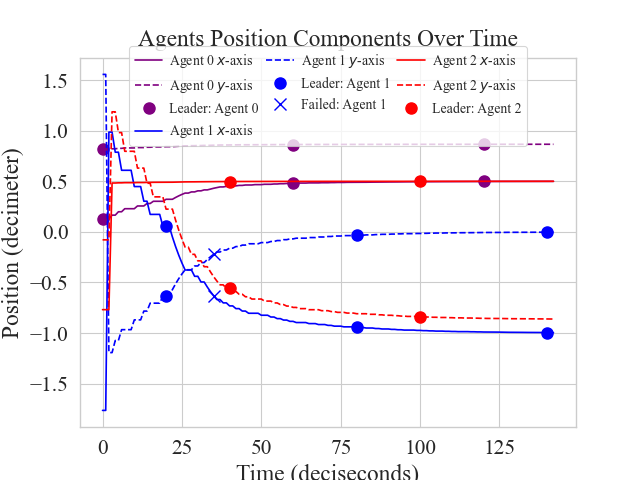}}
	\hfill
	\subfloat[d\label{6d}]{%
		\includegraphics[width=0.475\linewidth, trim={3mm 0mm 7mm 7mm}, clip]{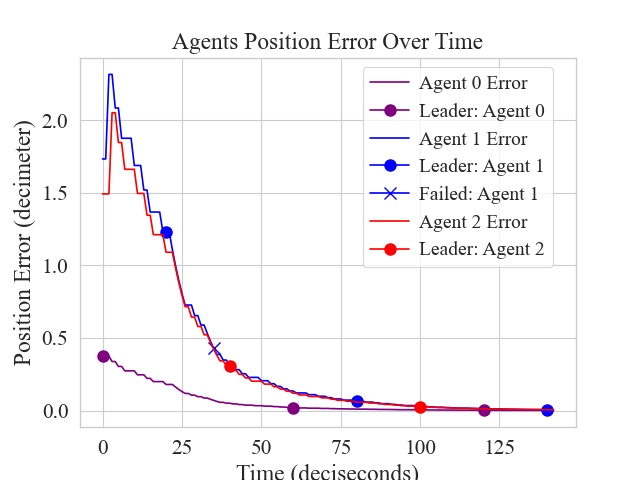}}
	\caption{Scenario D: Instead of updating the positions only on the leader node and then broadcasting the updated positions to the other nodes, the leader node can propose position updates and let the Raft algorithm replicate the new state across all nodes. (a): Agents' final position. (b): Agents trajectories from start to end. (c): Agents poistion components ($x$ and $y$) over time. (d): Agents position errors over time.}
	\label{fig6} 
\end{figure}

In Figure \ref{fig6}, by plotting the error between each agent's position and its goal position over time. This will allow to observe if there are any significant inconsistencies or jumps in the error when the leader changes or fails. Since the error plots show smooth transitions without any abrupt changes, this indicates that the state is consistent across all nodes, even during leader changes or failures.

In Scenario \ref{v6}, we replace the current leader selection process with Raft's leader election and introduce a randomized election timeout for each agent, which will trigger a new election if no leader is detected within that time. Finally, we ensure that agents vote for a leader and manage their votes according to the Raft algorithm.

\subsection{Scenario E}\label{sc_v7e}
%v7E
The agents have three possible statuses: follower, candidate, and leader. The leader is responsible for updating the positions of all agents to maintain the desired formation. If an agent does not receive a heartbeat from the leader for a certain duration, it starts an election and becomes a candidate.

%\begin{figure}[htbp] 
%    \centering
%  \subfloat[a\label{5a}]{%
%       \includegraphics[width=0.475\linewidth]{Imgs_f_a=n_v7/Figure_1.png}}
%    \hfill
%    \subfloat[b\label{5b}]{%
%    \includegraphics[width=0.475\linewidth]{Imgs_f_a=n_v7/Figure_2.png}}
%    \\
%  \subfloat[b\label{5c}]{%
%        \includegraphics[width=0.475\linewidth]{Imgs_f_a=n_v7/Figure_3.png}}
%    \hfill
%  \subfloat[c\label{5d}]{%
%        \includegraphics[width=0.475\linewidth]{Imgs_f_a=n_v7/Figure_4.png}}
%    \caption{Scenario E}
%  \label{fig5} 
%\end{figure}

\begin{figure}[htbp] 
	\centering
	\subfloat[a\label{7a}]{%
		\includegraphics[width=0.475\linewidth, trim={7mm 7mm 7mm 12mm}, clip]{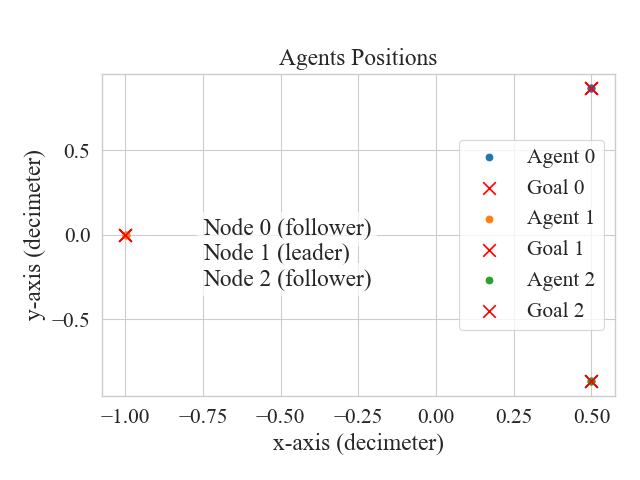}}
	\hfill
	\subfloat[b\label{7b}]{%
		\includegraphics[width=0.475\linewidth, trim={10mm 7mm 5mm 40mm}, clip]{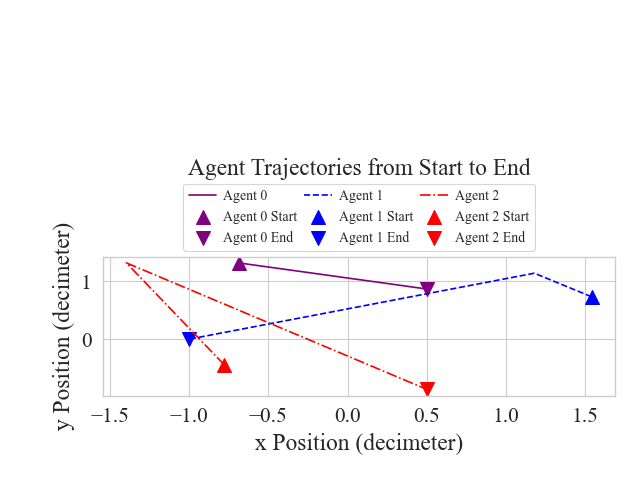}}
	\\
	\subfloat[b\label{7c}]{%
		\includegraphics[width=0.475\linewidth, trim={1mm 0mm 5mm 8mm}, clip]{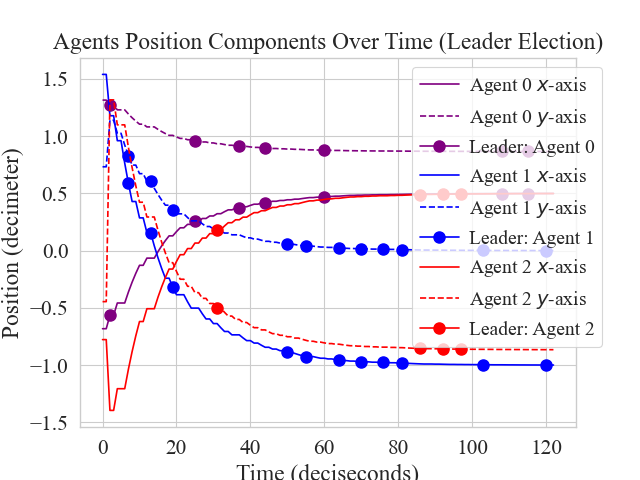}}
	\hfill
	\subfloat[c\label{7d}]{%
		\includegraphics[width=0.475\linewidth, trim={1mm 0mm 7mm 8mm}, clip]{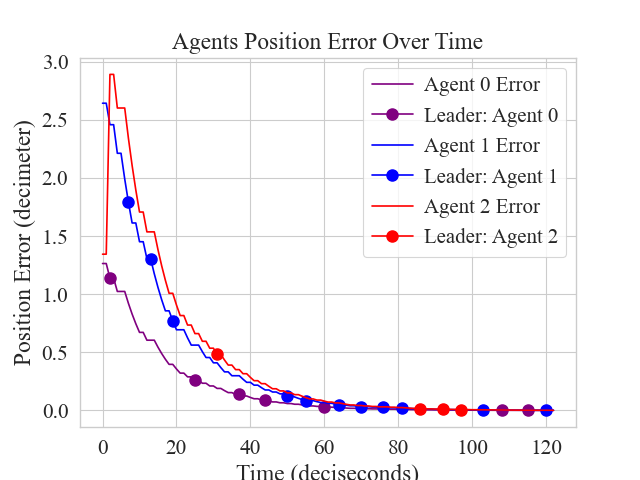}}
	\caption{Scenario E: (a): Agents' final position. (b): Agents trajectories from start to end. (c): Agents poistion components ($x$ and $y$) over time. (d): Agents position errors over time.}
	\label{fig7} 
\end{figure}

In Figure \ref{fig7}, each circle represents the position of an agent according to one of the nodes. Since you have three agents and three nodes, you get a total of 3 x 3 = 9 circles at the beginning.

\subsection{Scenario F}\label{sc_v8e}
\begin{figure}[htbp] 
    \centering
  \subfloat[a\label{8a}]{%
       \includegraphics[width=0.475\linewidth, trim={7mm 7mm 7mm 12mm}, clip]{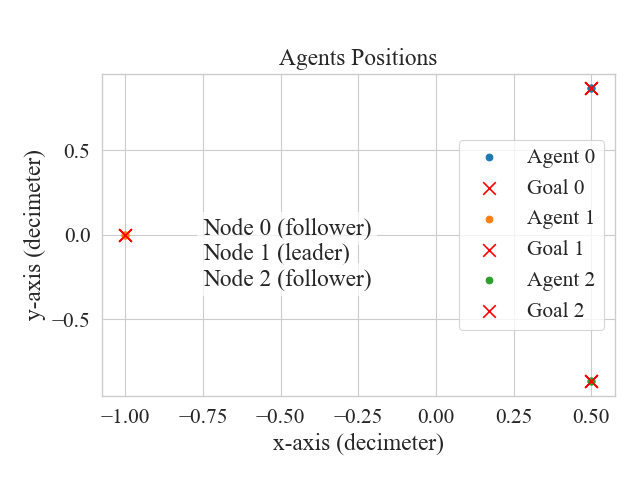}}
    \hfill
    \subfloat[b\label{8b}]{%
    \includegraphics[width=0.475\linewidth, trim={10mm 7mm 5mm 40mm}, clip]{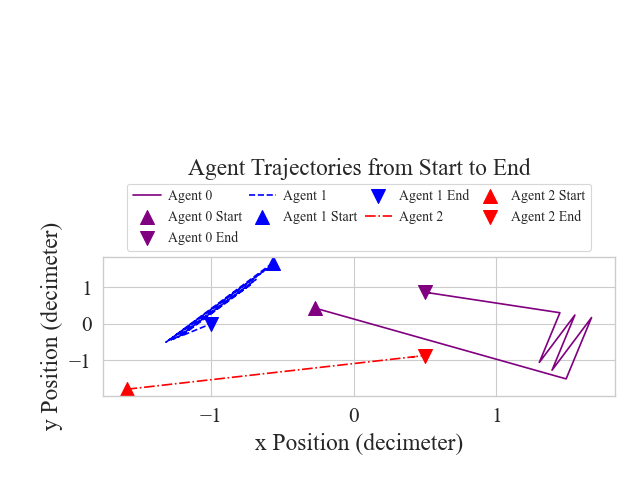}}
    \\
  \subfloat[b\label{8c}]{%
        \includegraphics[width=0.475\linewidth, trim={1mm 0mm 5mm 8mm}, clip]{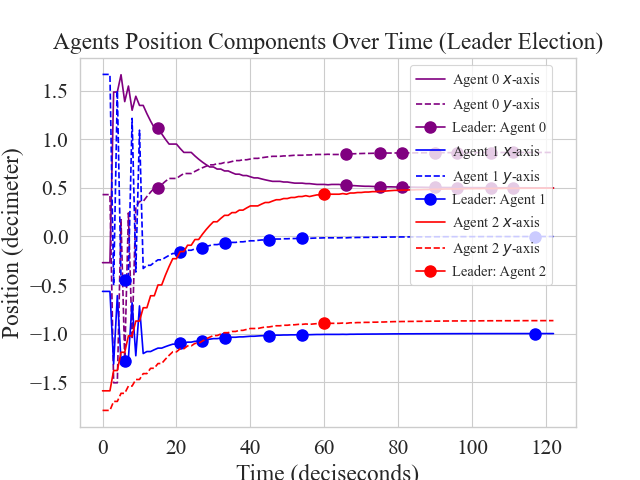}}
    \hfill
  \subfloat[c\label{8d}]{%
    \includegraphics[width=0.475\linewidth, trim={1mm 0mm 7mm 8mm}, clip]{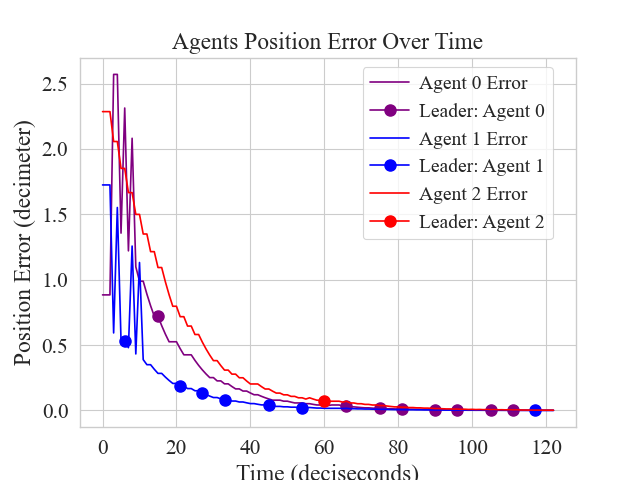}}
    \caption{Scenario F: Implementing a mechanism to detect and handle node failures, such as timeouts or unreachable nodes. (a): Agents' final position. (b): Agents trajectories from start to end. (c): Agents poistion components ($x$ and $y$) over time. (d): Agents position errors over time.}
  \label{fig8E} 
\end{figure}

%sc_v8E
Now we implement a mechanism to detect and handle node failures, such as timeouts or unreachable nodes. This method checks for failed nodes based on their status and the time passed since the last heartbeat received. If a follower node has not received a heartbeat for more than the specified timeout, it is marked as failed. To show the effictevness of the proposed method, we manually simulate node failures and recovery. In the simulation as shown in Figure \ref{fig8E}, we make node 1 to fail at frame 10 and recover at frame 20. 

\begin{table}
    \centering
    \caption{Status od nodes in different times and election terms in Scenario F}
    \label{tab:faultTolerant}
	\begin{tabular}{llrr}
		\toprule
		type              & node           &   term &   frame \\
		\midrule
		candidate         & 1              &      1 &       6 \\
		leader            & 1              &      1 &       6 \\
		simulate failure  & localhost:4322 &      1 &      10 \\
		failure           & localhost:4322 &      1 &      12 \\
		candidate         & 0              &      2 &      15 \\
		leader            & 0              &      2 &      15 \\
		failure           & localhost:4322 &      1 &      15 \\
		failure           & localhost:4322 &      1 &      18 \\
		simulate recovery & localhost:4322 &      1 &      20 \\
		candidate         & 1              &      2 &      21 \\
		leader            & 1              &      2 &      21 \\
		candidate         & 1              &      3 &      27 \\
		leader            & 1              &      3 &      27 \\
		candidate         & 1              &      4 &      33 \\
		leader            & 1              &      4 &      33 \\
		candidate         & 1              &      5 &      45 \\
		leader            & 1              &      5 &      45 \\
		candidate         & 1              &      6 &      54 \\
		leader            & 1              &      6 &      54 \\
		candidate         & 2              &      7 &      60 \\
		\bottomrule
	\end{tabular}
\end{table}

\subsection{Scenario G}\label{sc_v9EE}

\begin{figure}[htbp] 
  \centering
\subfloat[a\label{9a}]{%
     \includegraphics[width=0.475\linewidth, trim={7mm 7mm 7mm 12mm}, clip]{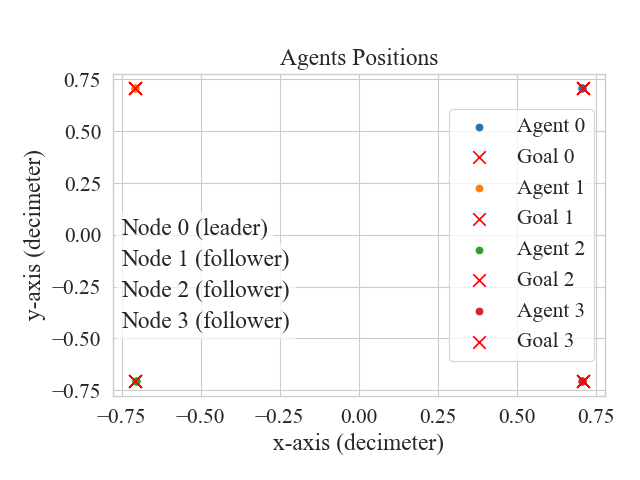}}
  \hfill
  \subfloat[b\label{9b}]{%
  \includegraphics[width=0.475\linewidth, trim={10mm 7mm 5mm 40mm}, clip]{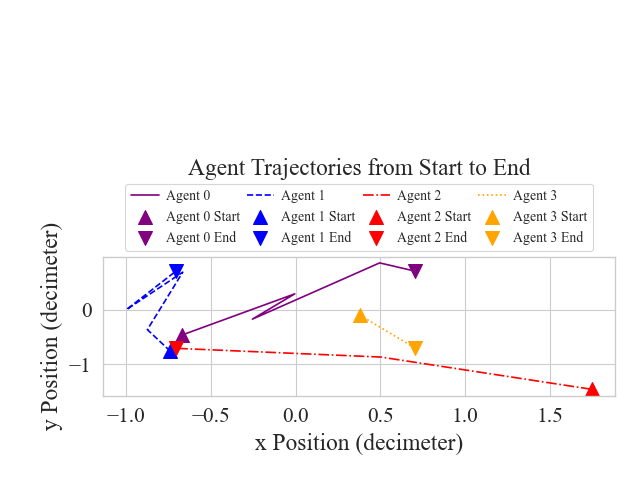}}
  \\
\subfloat[b\label{9c}]{%
      \includegraphics[width=0.475\linewidth, trim={1mm 0mm 5mm 8mm}, clip]{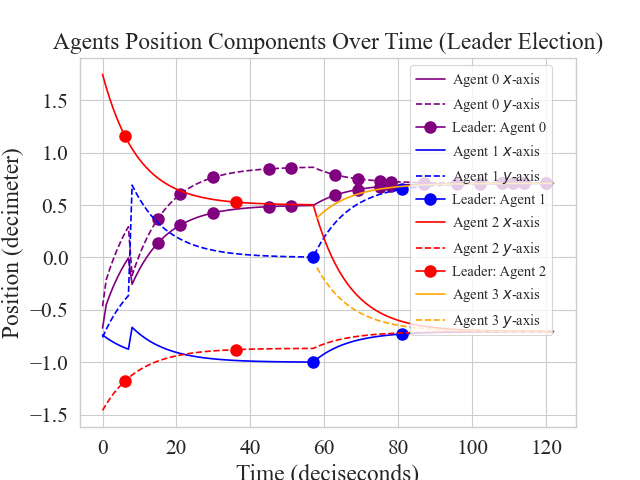}}
  \hfill
\subfloat[c\label{9d}]{%
  \includegraphics[width=0.475\linewidth, trim={1mm 0mm 7mm 8mm}, clip]{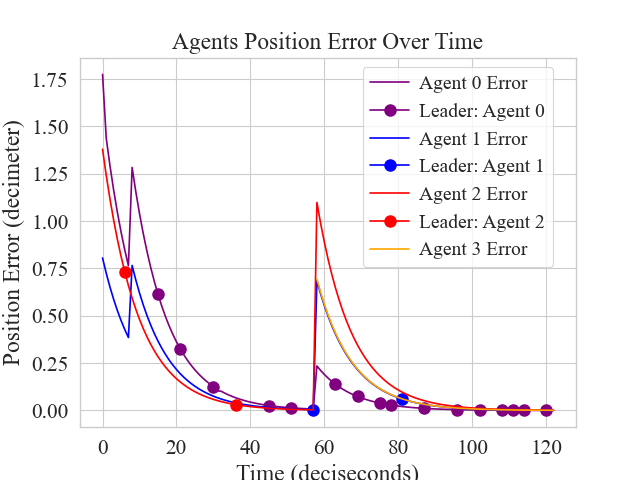}}
  \caption{Scenario G: Implementing dynamic membership changes, allowing to add or remove agents at runtime. (a): Agents' final position. (b): Agents trajectories from start to end. (c): Agents poistion components ($x$ and $y$) over time. (d): Agents position errors over time.}
\label{fig9} 
\end{figure}

In the last scenario that we consider here, the focus is on  implementing dynamic membership changes, allowing to add or remove agents at runtime. Raft supports membership changes through a joint consensus approach, which guarantees that the system remains operational during configuration changes. The agent or node removal is discussed in \ref{v5} and \ref{v5E}, therefore, we modify the approach and simulate a scenario that a node, i.e., an agent is added to the current cluster at runtime. The results are shown in Figure \ref{fig9}

\section{Discussion and Conclusions}
In this study, we address distributed formation control of multi-agent systems, combining leader election and node failures. The decentralized algorithm allows the system to function despite node failures or recoveries. The ElectionTimer class prevents simultaneous elections with a randomized timeout, while the DistributedFormation class manages agent positions and goals. Agents autonomously adjust their formation based on the leader's goals, which change with leader elections.

The formation control algorithm ensures agents form the desired shape, while the Raft algorithm provides a distributed, fault-tolerant method for maintaining consistent state across nodes. By synchronizing agents' states, the Raft algorithm allows the system to handle node failures while achieving desired formation control.

The implementation is robust against node failures, supports adding new agents during runtime, and allows for failed node recovery. However, the leader election process is not optimized for large-scale networks, potentially causing increased latency, and the simulation assumes a 2D environment.

%However, applying Raft to dynamical systems has challenges and limitations, such as latency, scalability, complexity, overhead, and use-case fit. Raft is not designed for real-time systems and may introduce latency, affecting fast response times. Its performance can degrade with increased node numbers, and implementing Raft adds complexity, potentially increasing development time and costs. Additionally, Raft requires communication and computational resources, which may not suit resource-constrained systems, and may not fit all robotic use cases.
%
%Future work could include optimizing the leader election process for large-scale systems, extending the algorithm to 3D environments, and incorporating different communication models for more realistic simulations. Additionally, further research could explore the impact of various network topologies and communication delays on the system's performance. Overall, this study provides a foundation for further exploration into distributed formation control and leader election algorithms in multi-agent systems.

However, while applying Raft to dynamical systems offers exciting possibilities, there are inherent challenges and limitations that should be addressed, such as latency, scalability, complexity, overhead, and use-case fit. Raft, for instance, is not initially designed for real-time systems, potentially introducing latency that could affect rapid response times. This is an area where future work could focus on optimizing the algorithm, specifically the leader election process, for quicker response times even in large-scale systems.

Moreover, Raft's performance may degrade with an increase in node numbers, and its implementation can add complexity, potentially increasing development time and costs. This identifies another opportunity for future enhancements, possibly through exploring different communication models for more realistic simulations and extending the algorithm's applicability to complex 3D environments.

Raft also necessitates communication and computational resources, which might not suit resource-constrained systems, and may not fit all robotic use cases. This calls for additional research to examine the impact of various network topologies and communication delays on the system’s performance.

In conclusion, the challenges with using Raft in dynamical systems are complex, but they give us a clear path for future work. This study reveals these challenges, yet it lays the groundwork for more research on distributed control and leader selection in systems with many agents. The limitations we've discussed aren't barriers, but signs showing us the way forward in this exciting area of study.

\bibliographystyle{ieeetr}
\bibliography{refs}

\begin{thebibliography}{10}

\bibitem{wooldridge2009introduction}
M.~Wooldridge, {\em An introduction to multi-agent systems}.
\newblock John Wiley \& Sons, 2009.

\bibitem{coulouris2005distributed}
G.~Coulouris, J.~Dollimore, and T.~Kindberg, {\em Distributed Systems: Concepts
  and Design}.
\newblock Addison-Wesley, 2005.

\bibitem{olfati2004consensus}
R.~Olfati-Saber and R.~M. Murray, ``Consensus problems in networks of agents
  with switching topology and time-delays,'' {\em IEEE Transactions on
  Automatic Control}, vol.~49, no.~9, pp.~1520--1533, 2004.

\bibitem{kempe2003gossip}
D.~Kempe, A.~Dobra, and J.~Gehrke, ``Gossip-based computation of aggregate
  information,'' in {\em 44th Annual IEEE Symposium on Foundations of Computer
  Science, 2003. Proceedings}, pp.~482--491, IEEE, 2003.

\bibitem{cao2013overview}
Y.~Cao, W.~Yu, W.~Ren, and G.~Chen, ``An overview of recent progress in the
  study of distributed multi-agent coordination,'' {\em IEEE Transactions on
  Industrial Informatics}, vol.~9, no.~1, pp.~427--438, 2013.

\bibitem{stone2000multi}
P.~Stone and M.~Veloso, ``Multi-agent systems: A survey from a machine learning
  perspective,'' {\em Autonomous Robots}, vol.~8, no.~3, pp.~345--383, 2000.

\bibitem{ongaro2014search}
D.~Ongaro and J.~Ousterhout, ``In search of an understandable consensus
  algorithm,'' in {\em 2014 {USENIX} Annual Technical Conference ({USENIX}
  {ATC} 14)}, pp.~305--319, USENIX Association, 2014.

\bibitem{lamport1998part}
L.~Lamport, ``The part-time parliament,'' {\em ACM Transactions on Computer
  Systems}, vol.~16, no.~2, pp.~133--169, 1998.

\bibitem{howard2014raft}
H.~Howard, M.~Schwarzkopf, A.~Madhavapeddy, and S.~Hand, ``Raft refloated: Do
  we have consensus?,'' {\em ACM SIGOPS Operating Systems Review}, vol.~49,
  no.~1, pp.~12--21, 2014.

\bibitem{moraru2013more}
I.~Moraru, D.~G. Andersen, and M.~Kaminsky, ``There is more consensus in
  egalitarian parliaments,'' in {\em Proceedings of the 24th ACM Symposium on
  Operating Systems Principles}, pp.~358--372, ACM, 2013.

\bibitem{jin2018optimized}
L.~Jin, C.~Xu, Y.~Deng, and Q.~Zhang, ``An optimized consensus algorithm based
  on the raft protocol,'' {\em Concurrency and Computation: Practice and
  Experience}, vol.~30, no.~2, 2018.

\bibitem{sundaravel2018raft}
S.~Sundaravel, D.~M. Divakaran, and O.~W. Hock, ``Raft-based consensus for
  large-scale iot networks,'' in {\em 2018 15th IEEE Annual Consumer
  Communications \& Networking Conference (CCNC)}, pp.~1--4, IEEE, 2018.

\bibitem{pramanik2019blockraft}
R.~Pramanik, R.~Mall, and M.~Chattopadhyay, ``Blockraft: A blockchain consensus
  algorithm using the raft consensus protocol,'' {\em Journal of Parallel and
  Distributed Computing}, vol.~129, pp.~107--120, 2019.

\bibitem{zhang2017raftsdn}
Y.~Zhang, S.~Guo, Y.~Liu, and C.~Hu, ``Raftsdn: A consistent and fault-tolerant
  sdn controller based on raft algorithm,'' {\em Computer Communications},
  vol.~111, pp.~1--10, 2017.

\bibitem{sharma2018craq}
N.~Sharma, J.~Fonseca, and M.~S. Ardekani, ``Craq: Consistent, scalable, and
  crash-resilient replicated lists using the raft consensus algorithm,'' {\em
  ACM Transactions on Storage (TOS)}, vol.~14, no.~3, pp.~1--33, 2018.

\bibitem{ardekani2014samsara}
M.~S. Ardekani, N.~Sharma, and J.~Fonseca, ``Samsara: Efficient deterministic
  replay in large-scale data processing systems,'' in {\em 2014 ACM/IFIP/USENIX
  Middleware}, pp.~207--218, ACM, 2014.

\bibitem{yang2019raft}
H.~Yang and A.~Y. Zomaya, ``Raft-based edge computing for mobile iot,'' {\em
  IEEE Internet of Things Journal}, vol.~6, no.~3, pp.~4877--4885, 2019.

\bibitem{medeiros2018vc}
S.~A. Medeiros, R.~C. Moraes, F.~C. Delicato, E.~Cavalcante, and P.~F. Pires,
  ``Vc-raft: A raft-based vehicular cloud architecture for platooning,'' in
  {\em 2018 17th IEEE International Symposium on Network Computing and
  Applications (NCA)}, pp.~1--8, IEEE, 2018.

\bibitem{liu2020raft}
R.~Liu, J.~Luo, S.~Shen, Y.~Cheng, and Y.~Gu, ``Raft-based consensus for
  cooperation among uav swarms,'' in {\em 2020 IEEE 6th International
  Conference on Computer and Communications (ICCC)}, pp.~2295--2299, IEEE,
  2020.

\bibitem{ren2020distributed}
W.~Ren, J.~Huang, and Y.~Zhao, ``Distributed formation control using observers
  and output feedback,'' {\em Automatica}, vol.~117, p.~108948, 2020.

\bibitem{oh2019leader}
K.~W. Oh, C.~Park, and J.~Lee, ``Leader-follower formation control for
  multi-agent systems with second-order dynamics,'' {\em International Journal
  of Control, Automation and Systems}, vol.~17, no.~11, pp.~2797--2808, 2019.

\bibitem{cao2017consensus}
Y.~Cao, Y.~Hong, and W.~Ren, ``Consensus in high-dimensional multi-agent
  systems,'' {\em IEEE Transactions on Automatic Control}, vol.~62, no.~12,
  pp.~6399--6404, 2017.

\bibitem{li2018distributed}
S.~Li, Y.~Peng, and H.~Wang, ``Distributed formation control for uav swarm
  systems with communication delays,'' {\em International Journal of Control,
  Automation and Systems}, vol.~16, no.~6, pp.~2821--2828, 2018.

\bibitem{liu2017consensus}
H.~Liu, G.~Chen, E.~Blasch, and K.~Pham, ``A consensus-based approach for
  tracking a maneuvering target using a swarm of quadrotors,'' {\em IEEE
  Access}, vol.~5, pp.~24443--24452, 2017.

\bibitem{zhu2018distributed}
J.~Zhu and G.~Hug, ``Distributed consensus algorithm for multi-agent systems in
  smart grid applications,'' {\em IEEE Transactions on Smart Grid}, vol.~9,
  no.~5, pp.~5226--5236, 2018.

\bibitem{chen2018formation}
G.~Chen and Q.~Meng, ``Formation control for mobile sensor networks: A
  consensus-based approach,'' {\em ISA Transactions}, vol.~74, pp.~139--148,
  2018.

\bibitem{wang2019formation}
W.~Wang, J.-Q. Leng, and J.~Wu, ``Formation control for cooperative adaptive
  cruise control,'' {\em IEEE Transactions on Intelligent Transportation
  Systems}, vol.~20, no.~7, pp.~2633--2647, 2019.

\bibitem{sun2019distributed}
Z.~Sun, G.~Ma, B.~Jiao, M.~Zhou, and F.~Yan, ``Distributed consensus algorithm
  for cooperative localization in connected vehicle systems,'' {\em IEEE
  Transactions on Vehicular Technology}, vol.~68, no.~5, pp.~4185--4198, 2019.

\bibitem{zhang2018consensus}
C.~Zhang, Y.~Lv, and L.~Pan, ``A consensus-based approach for community
  detection in social networks,'' {\em Physica A: Statistical Mechanics and its
  Applications}, vol.~508, pp.~1--12, 2018.

\end{thebibliography}


\begin{thebibliography}{10}

\bibitem{wooldridge2009introduction}
M.~Wooldridge, {\em An introduction to multi-agent systems}.
\newblock John Wiley \& Sons, 2009.

\bibitem{coulouris2005distributed}
G.~Coulouris, J.~Dollimore, and T.~Kindberg, {\em Distributed Systems: Concepts
  and Design}.
\newblock Addison-Wesley, 2005.

\bibitem{hu2020distributed}
J.~Hu, {\em Distributed Formation Control of Multi-Agent Systems: Theory and
  Applications}.
\newblock The University of Manchester (United Kingdom), 2020.

\bibitem{kempe2003gossip}
D.~Kempe, A.~Dobra, and J.~Gehrke, ``Gossip-based computation of aggregate
  information,'' in {\em 44th Annual IEEE Symposium on Foundations of Computer
  Science, 2003. Proceedings}, pp.~482--491, IEEE, 2003.

\bibitem{brambilla2013swarm}
M.~Brambilla, E.~Ferrante, M.~Birattari, and M.~Dorigo, ``Swarm robotics: a
  review from the swarm engineering perspective,'' {\em Swarm Intelligence},
  vol.~7, pp.~1--41, 2013.

\bibitem{he2016distributed}
L.~He, X.~Sun, and Y.~Lin, ``Distributed output-feedback formation tracking
  control for unmanned aerial vehicles,'' {\em International Journal of Systems
  Science}, vol.~47, no.~16, pp.~3919--3928, 2016.

\bibitem{martinson2005lattice}
E.~Martinson and D.~Payton, ``Lattice formation in mobile autonomous sensor
  arrays,'' in {\em Swarm Robotics: SAB 2004 International Workshop, Santa
  Monica, CA, USA, July 17, 2004, Revised Selected Papers 1}, pp.~98--111,
  Springer, 2005.

\bibitem{cao2013overview}
Y.~Cao, W.~Yu, W.~Ren, and G.~Chen, ``An overview of recent progress in the
  study of distributed multi-agent coordination,'' {\em IEEE Transactions on
  Industrial Informatics}, vol.~9, no.~1, pp.~427--438, 2013.

\bibitem{stone2000multi}
P.~Stone and M.~Veloso, ``Multi-agent systems: A survey from a machine learning
  perspective,'' {\em Autonomous Robots}, vol.~8, no.~3, pp.~345--383, 2000.

\bibitem{ongaro2014search}
D.~Ongaro and J.~Ousterhout, ``In search of an understandable consensus
  algorithm,'' in {\em 2014 {USENIX} Annual Technical Conference ({USENIX}
  {ATC} 14)}, pp.~305--319, USENIX Association, 2014.

\bibitem{lamport1998part}
L.~Lamport, ``The part-time parliament,'' {\em ACM Transactions on Computer
  Systems}, vol.~16, no.~2, pp.~133--169, 1998.

\bibitem{howard2014raft}
H.~Howard, M.~Schwarzkopf, A.~Madhavapeddy, and S.~Hand, ``Raft refloated: Do
  we have consensus?,'' {\em ACM SIGOPS Operating Systems Review}, vol.~49,
  no.~1, pp.~12--21, 2014.

\bibitem{moraru2013more}
I.~Moraru, D.~G. Andersen, and M.~Kaminsky, ``There is more consensus in
  egalitarian parliaments,'' in {\em Proceedings of the 24th ACM Symposium on
  Operating Systems Principles}, pp.~358--372, ACM, 2013.

\bibitem{olfati2004consensus}
R.~Olfati-Saber and R.~M. Murray, ``Consensus problems in networks of agents
  with switching topology and time-delays,'' {\em IEEE Transactions on
  Automatic Control}, vol.~49, no.~9, pp.~1520--1533, 2004.

\bibitem{fu2021improved}
W.~Fu, X.~Wei, and S.~Tong, ``An improved blockchain consensus algorithm based
  on raft,'' {\em Arabian Journal for Science and Engineering}, vol.~46, no.~9,
  pp.~8137--8149, 2021.

\bibitem{liu2021dqn}
Z.~Liu, L.~Hou, K.~Zheng, Q.~Zhou, and S.~Mao, ``A dqn-based consensus
  mechanism for blockchain in iot networks,'' {\em IEEE Internet of Things
  Journal}, vol.~9, no.~14, pp.~11962--11973, 2021.

\bibitem{wang2019k}
R.~Wang, L.~Zhang, Q.~Xu, and H.~Zhou, ``K-bucket based raft-like consensus
  algorithm for permissioned blockchain,'' in {\em 2019 IEEE 25th International
  Conference on Parallel and Distributed Systems (ICPADS)}, pp.~996--999, IEEE,
  2019.

\bibitem{gonzalez2016fault}
A.~J. Gonzalez, G.~Nencioni, B.~E. Helvik, and A.~Kamisinski, ``A
  fault-tolerant and consistent sdn controller,'' in {\em 2016 IEEE global
  communications conference (GLOBECOM)}, pp.~1--6, IEEE, 2016.

\bibitem{deyerl2019search}
C.~Deyerl and T.~Distler, ``In search of a scalable raft-based replication
  architecture,'' in {\em Proceedings of the 6th Workshop on Principles and
  Practice of Consistency for Distributed Data}, pp.~1--7, 2019.

\bibitem{hou2021intelligent}
L.~Hou, X.~Xu, K.~Zheng, and X.~Wang, ``An intelligent transaction migration
  scheme for raft-based private blockchain in internet of things
  applications,'' {\em IEEE Communications Letters}, vol.~25, no.~8,
  pp.~2753--2757, 2021.

\bibitem{nasimi2020platoon}
M.~Nasimi, M.~A. Habibi, and H.~D. Schotten, ``Platoon--assisted vehicular
  cloud in vanet: Vision and challenges,'' {\em arXiv preprint
  arXiv:2008.10928}, 2020.

\bibitem{cheng2019distributed}
B.~Cheng and Z.~Li, ``Distributed formation control via output feedback
  event-triggered coordination,'' in {\em 2019 Chinese Control And Decision
  Conference (CCDC)}, pp.~880--885, IEEE, 2019.

\bibitem{li2013leader}
W.~Li, Z.~Chen, and Z.~Liu, ``Leader-following formation control for
  second-order multiagent systems with time-varying delay and nonlinear
  dynamics,'' {\em Nonlinear Dynamics}, vol.~72, pp.~803--812, 2013.

\bibitem{xiao2007consensus}
F.~Xiao and L.~Wang, ``Consensus problems for high-dimensional multi-agent
  systems,'' {\em IET Control Theory \& Applications}, vol.~1, no.~3,
  pp.~830--837, 2007.

\bibitem{kartal2020distributed}
Y.~Kartal, K.~Subbarao, N.~R. Gans, A.~Dogan, and F.~Lewis, ``Distributed
  backstepping based control of multiple uav formation flight subject to time
  delays,'' {\em IET Control Theory \& Applications}, vol.~14, no.~12,
  pp.~1628--1638, 2020.

\bibitem{nair2018multi}
A.~S. Nair, T.~Hossen, M.~Campion, D.~F. Selvaraj, N.~Goveas, N.~Kaabouch, and
  P.~Ranganathan, ``Multi-agent systems for resource allocation and scheduling
  in a smart grid,'' {\em Technology and Economics of Smart Grids and
  Sustainable Energy}, vol.~3, pp.~1--15, 2018.

\bibitem{qian2023formation}
X.~Qian and B.~Cui, ``Formation control of large-scale mobile sensor networks
  based on semilinear parabolic system,'' {\em The Journal of Engineering},
  vol.~2023, no.~1, p.~e12215, 2023.

\bibitem{wang2018review}
Z.~Wang, G.~Wu, and M.~J. Barth, ``A review on cooperative adaptive cruise
  control (cacc) systems: Architectures, controls, and applications,'' in {\em
  2018 21st International Conference on Intelligent Transportation Systems
  (ITSC)}, pp.~2884--2891, IEEE, 2018.

\bibitem{liu2016cooperative}
J.~Liu, B.-g. Cai, and J.~Wang, ``Cooperative localization of connected
  vehicles: Integrating gnss with dsrc using a robust cubature kalman filter,''
  {\em IEEE Transactions on Intelligent Transportation Systems}, vol.~18,
  no.~8, pp.~2111--2125, 2016.

\bibitem{karimi2018consensus}
A.-M. Karimi-Majd, M.~Fathian, and M.~Makrehchi, ``Consensus-based methodology
  for detection communities in multilayered networks,'' {\em Physica A:
  Statistical Mechanics and its Applications}, vol.~494, pp.~547--558, 2018.

\end{thebibliography}

\end{document}